# Platform-Independent and Curriculum-Oriented Intelligent Assistant for Higher Education

Ramteja Sajja[1,3], Yusuf Sermet[3], David Cwiertny[2,3,4,5], Ibrahim Demir[1,2,3]

[1] Department of Electrical Computer Engineering, University of Iowa
[2] Department of Civil and Environmental Engineering, University of Iowa
[3] IIHR Hydroscience and Engineering, University of Iowa
[4] Department of Chemistry, University of Iowa
[5] Center for Health Effects of Environmental Contamination, University of Iowa

**Abstract**
Miscommunication and communication challenges between instructors and students represents one of the primary barriers to post-secondary learning. Students often avoid or miss opportunities to ask questions during office hours due to insecurities or scheduling conflicts. Moreover, students need to work at their own pace to have the freedom and time for the self-contemplation needed to build conceptual understanding and develop creative thinking skills. To eliminate barriers to student engagement, academic institutions need to redefine their fundamental approach to education by proposing flexible educational pathways that recognize continuous learning. To this end, we developed an AI-augmented intelligent educational assistance framework based on a power language model (i.e., GPT-3) that automatically generates course-specific intelligent assistants regardless of discipline or academic level. The virtual intelligent teaching assistant (TA) system will serve as a voice-enabled helper capable of answering course-specific questions concerning curriculum, logistics and course policies. It is envisioned to improve access to course-related information for the students and reduce logistical workload for the instructors and TAs. Its GPT-3-based knowledge discovery component as well as the generalized system architecture is presented accompanied by a methodical evaluation of the system accuracy and performance.

**Keywords:** Artificial Intelligence, Natural Language Processing, Machine Learning, Transformers, GPT-3

## 1. Introduction

One of the main causes of the knowledge disparities that lead to learning gaps among both undergraduate and graduate students is instructors' inability to communicate with these students in ways that suit the students' learning schedules and styles (Williamson et al., 2020). It has been widely shown in the literature that it is particularly effective to teach in ways that allow students to build conceptual understanding of the subject they are studying (Konicek-Moran and Keeley, 2015). This requires a certain degree of freedom and time for self-contemplation (Lin and Chan, 2018). Not surprisingly, allowing students to learn at their own pace positively contributes to a substantial increase in learning motivation and the development of creative thinking skills (Ciampa, 2014).

A significant portion of students avoid or miss the opportunity to visit teaching assistants and instructors during office hours due to scheduling conflicts, the feeling of not being prepared, the imposter syndrome, and shyness (Abdul-Wahab et al., 2019). Furthermore, most students study outside of regular work hours, which creates needs for assistance at odd times (Mounsey et al., 2013). The lack of immediate assistance can lead to discouragement and creates the feeling of being stuck despite the fact that many queries can be simply answered based on available material without in-depth expertise (Seeroo and Bekaroo, 2021). Teaching assistants can sometimes fill this void, but they have their own responsibilities (e.g., classes, research, grading) which may render them unavailable during times such as exam weeks when the students need them most (Howitz et al., 2020). Thus, it would be extraordinarily helpful to develop new and more readily available forms of student assistance if this can be done without decreasing the time TAs and instructors have to spend on higher-level instruction (Mirzajani et al., 2016).

Information and communication tools and services are critical parts of instructional technology and the learning process. Web technologies support instructors in the delivery of curriculum on advanced modeling and analysis tools (Ewing et al., 2022), programming libraries (Ramirez et al., 2022), and engineering ethics using serious games (Ewing and Demir, 2021). AI has been used actively in two core areas, including information processing and knowledge communication. Deep learning models are commonly used for image processing (Li and Demir, 2023), data augmentation (Demiray et al., 2021), synthetic data generation (Gautam et al., 2022), and modeling studies (Sit et al., 2021). AI use cases for information communication and delivery are relatively limited in the engineering domain (Yesilkoy et al., 2022). Smart assistants based on customized ontologies (Sermet and Demir, 2019) have been actively used in public health care (Sermet et al., 2021) and environmental science (Sermet and Demir, 2018) studies.

With the recent advancements (i.e., ChatGPT) in AI based communication, there is a significant interest in the research of chatbots, which can be defined as intelligent agents (i.e., assistants) that have the ability to comprehend natural language queries and produce a direct and factual response utilizing data and service providers (Brandtzaeg and Følstad, 2017). Voice-based assistants are actively used in education, environmental science, and operational systems to access real-time data, train first responders (Sermet and Demir, 2022), and facilitate decision support coupled with other communication technologies like virtual and augmented reality (Sermet and Demir, 2020).

Technology companies have been taking the lead on operational virtual assistants integrated into their ecosystem which triggered a brand new and massive market that is forecasted to reach US$ 11.3 billion by 2024 (IMARC Group, 2019). Several studies emphasize the potential chatbots hold to serve as the next generation information communication tool and make the case for an urgent need for chatbot development and adoption in their respective fields (Androutsopoulou et al., 2019; Vaidyam et al., 2019; USACE, 2019; Miner et al., 2020). However, the usage of chatbots for effective and reliable information communication is not

widespread among public, government, scientific communities, and universities (Schoemaker and Tetlock, 2017). The adoption of virtual assistants within the context of academic curriculum can help closing the learning gaps identified above and, in the literature, (Hwang and Chang, 2021). Considering the prevalence of mobile phones and computers among students along with the recent remote-interaction culture that is gained during the pandemic, such technological and web-based solutions are relevant and needed more than ever (Iglesias-Pradas et al., 2021).

A recent report on the AI Market in the US Education Sector (TechNavio, 2018) emphasizes AI's focus on creating intelligent systems, discusses its increasing use in enhancing student learning, and states that intelligent interactive programs that are based on Machine Learning and Natural Language Processing help in overall learning of students. It is reported that the most significant market trend is the increased emphasis on chatbots (MindCommerce, 2019). The main aspect of how AI can be a vital tool in education is the utilization of AI in developing next-generation educational tools and solutions to provide a modern learning experience with the vision of personalized teaching, advising, and support (GATech, 2018; Ceha et al., 2021).

We propose an AI-augmented intelligent educational assistance framework that automatically generates course-specific intelligent assistants based on provided documents (e.g., syllabus) regardless of discipline or academic level. It will serve as a message-enabled helper capable of answering course-specific questions concerning scope and logistics (e.g., syllabus, deadlines, policies). The students can converse with the assistant in natural language via web platforms as well as messaging applications. The framework is conceived to address the listed issues and to unlock the immense potential of conversational AI approaches for education and enhancing the learning experience. Core benefits and advantages of the framework include the availability of assistance regardless of time, more TA and instructor time for advanced and customized advising, answers to time consuming and repetitive questions, reduced human error due to miscommunication for course logistics, and accommodations for personal barriers, cultural, and disability related issues (e.g., language barrier). A case study is conducted to quantitatively measure the proposed approach's efficacy and reliability within the context of the cited benefits.

The remainder of this article is organized as follows. Section 1.1 summarizes the relevant literature and identifies the knowledge gap. Section 2 presents the methodology of the design choices, development and implementation of a course-oriented intelligent assistance system based on syllabi. Section 3 describes the preliminary results and provides benchmark results and performance analysis. Section 4 concludes the articles with a summary of contributions and future work.

## 1.1. Related Work

There have been several initiatives to leverage conversational interfaces in education systems and higher learning (Hobert, 2019; Wollny et al., 2021; Chuaphan et al., 2021). Georgia Tech pioneered a virtual teaching assistant (TA) named "Jill Watson" and reported inspiring results for student satisfaction (GATech, 2018). Additionally, many students were inspired and created their own chatbots that converse about the courses, exhibiting increased interest in AI tools. The positive impacts of cultivating a teaching motivation for individual learning are successfully demonstrated in University of Waterloo's Curiosity Notebook research project, in which the students reported increased engagement upon conversation with an intelligent agent (i.e., Sigma) that asks Geology-related questions in a humorous manner (Ceha et al., 2021). Several universities have similar projects exploring AI's role in education including Stanford University (Ruan et al., 2019) and Carnegie Mellon University (Helmer, 2019). Further initiatives have been explored to utilize chatbots in certain aspects of campus life (Dibitonto et al., 2018; Duberry and Hamidi, 2021). Georgia State University developed a virtual assistant for admission support (i.e., Pounce) to incoming freshmen students. The Randomized Control Trial they implemented to assess effectiveness yielded that first-generation and underrepresented groups disproportionately benefited from the system which resulted in decreased gap in graduation rate among different demographics (Hart, 2019). Furthermore, 94% of the students recommended GSU to continue the service, citing their satisfaction in receiving instant responses any time of the day without the feeling of being judged or perceived as unintelligent (Mainstay, 2021).

The process of creating a knowledge framework includes retrieving relevant documents and extracting answers from the retrieved documents (Zylich et al., 2020). One of the main documents that can be used to acquire course information to answer logistical questions is syllabus (Zoroayka, 2018; Zylich et al., 2020). Chatbots can also be extended to other works such as helping students with technical issues and questions (Chuaphan et al., 2021). The limits of TA's human resources can be addressed with the help of these chatbots. A chatbot was deployed at Stanford University to respond to student inquiries for a class by compiling information from their participation in an online forum (Chopra et al., 2016). Similarly, a solution to augment the staffing shortages is with an AI Teaching Assistant (Ho, 2018). In addition to assisting with staff shortages, the virtual teaching assistants improve students' educational experiences (du Boulay, 2016). The chatbots can be developed internally using readily available open source tools (Zylich et al., 2020) or through the use of cloud-based language models (Benedetto et al., 2019; Chuaphan et al., 2021; Ranavare & Kamath, 2020).

The literature review clearly shows the importance of chatbots and how they can be used in the educational setting. Based on the survey done by Abdelhamid & Katz, 2020, more than 75% of students who responded to the study said they have previously used a chatbot service or other comparable system. 71% of the students stated that they find it challenging to meet with their teaching assistants for a variety of reasons. More than 95% of students claimed that having a chatbot available would be beneficial for providing some of their questions with answers. Though the previous work puts forth limited-scope case studies that clearly serve as proof of potential and benefits of utilizing conversational approaches in the educational setting, a complete and multidisciplinary solution has not been introduced to transform teaching and learning. A major distinction of the proposed framework in contrast to relevant work is the ability to automatically generate a ready-to-use intelligent assistant based on dynamic input provided in the format of a textual document, such as a curriculum summary and syllabus. It is both independent from the field and the technology (e.g., learning management systems) used for the content delivery. Furthermore, it relies on a Service-Oriented Architecture (SOA) to enable integration to any delivery channel.

## 2. Methodology
### 2.1. Natural Language Inference

In recent years, the rapid expansion of data volume was facilitated by technological innovation. A Forbes survey from a few years back revealed that 2.5 quintillion bytes of data were produced every single day. According to current estimates, unstructured data makes up more than 90% of all information stored (Kim et al., 2014). The introduction of language models as a foundation for numerous applications trying to extract useful insights from unstructured text was one of the major forces behind such research. In order to anticipate words, a language model analyzes the structure of human language. There are multiple available large language models such as BERT (Devlin et al., 2019), XLNet (Yang et al., 2020), RoBERTa (Liu et al., 2019), ALBERT (Lan et al., 2020), GPT-3 (Brown et al., 2020), GPT-2 (Radford et al., 2019), and PaLM (Chowdhery et al., 2022).

OpenAI provides the generative pre-trained Transformer 3 (GPT-3), an autoregressive language model that uses deep learning to generate writing that resembles that of a human. GPT-3 can be utilized off the shelf as well as by using a few-shot learning technique and fine-tuning the model in order to adapt it to any desired application area.GPT-3 has been pre-trained on a large quantity of text from the public internet resources, which may be regarded as few-shot learning. When given only a few instances, it can typically figure out what task you're attempting to complete and offer a convincing solution. Fine-tuning builds on fine-tuning learning by training on many more instances that can fit in the prompt, allowing to obtain higher outcomes. Once a model has been fine-tuned, it no longer needs examples in the prompt. Fine tuning also reduces expenses and makes lower-latency requests possible. We decided to choose GPT-3 because the model is cloud based and has a developer friendly API. GPT-3's Davinci is the biggest model in terms of parameters that's available to use by researchers and general public.

### 2.2. Syllabus Knowledge Model

It is crucial to pick the correct questions to pose to the chatbot in order to test its accuracy. The key questions that could arise from the course syllabus were extracted using the literature on syllabus templates because the goal of this research was to create a chatbot that could answer questions using the course logistics and policies from syllabus. The important sections of a syllabus or course description is Course Information, Faculty Information, Instructional Materials and Methods, Learning Outcomes, Grading and Course Expectations, Policies, Course Schedule (Hess et al., 2007; Passman & Green, 2009). Similarly critical sections such as disability statements, academic misconduct, inclusivity, accessibility and harassment; and optional information such as mental health resources can also be included in the syllabus (Wagner et al., 2022). Based on this literature, we included questions that were related to the following topics: (1) Course Information, (2) Faculty Information, (3) Teaching Assistant Information, (4) Course Goals, (5) Course Calendar, (6) Attendance, (7) Grading, (8) Instructional Materials, and (9) Course and Academic Policies. Course Information, Faculty Information and TA Information sections of the knowledge graph developed to encompass a standard syllabus in higher education is shown in Table 1. After analyzing all the main categories specified above, we have generated 36 questions to reflect information included in these categories. We also used text and data augmentation techniques on these 36 questions to generate 120 questions in total to reflect different ways a question could be asked. This text and data augmentation approaches are reflected in Table A.1 in the form of competency questions.

Table 1. Knowledge Graph for a Syllabus in Higher Education

| Category | Syllabus Element |
|---|---|
| Course Information | Course Name |
| | Course Number |
| | Credit Hours |
| | Location and Class Times |
| Faculty Information | Name |
| | Contact Information |
| | Office Location |
| | Office Hours |
| TA/Teaching Assistant Information | Name |
| | Contact Information |
| | Office Location |
| | Office Hours |
| Course Goals and/or Objectives | Course Objectives |
| | Expectations from the course |
| Course Calendar | Due dates |
| | Assignment dates |
| Attendance and classroom behavior | Attendance policy |
| | Expected Classroom behavior |
| Grading and Assignments | Grading Criteria |
| | Tentative Exam Schedule |
| Instructional Materials | Textbooks |
| | Other required materials for the course |
| Policies | Late Assignments |
| | Academic Dishonesty |
| | Disability Statement |
| | Freedom of Speech |
| | Makeup Policy |
| | Mental Health Resources |
| | Absences for Religious Holy Days |

## 2.3. System Architecture

The VirtualTA framework can be partitioned into four major cyber components with specialized functions (Figure 1). The first component can be attributed to curating and indexing appropriate classroom resources for information that falls within the scope of a syllabus, as described in Table 1. The second component contains the cyber framework to create, serve, and manage the smart assistant. It includes server management, API access from the perspectives of both students and instructors, data analytics, and smart assistant management. Intelligent Service component is concerned about the deep learning-powered natural language tools that are provided under the umbrella of VirtualTA framework (e.g., inference and intent mapping, emotion detection and lightening the mood with witty yet helpful responses). Finally, the integration component is concerned with communication channels the smart assistant can be served from and entails the appropriate protocols, webhooks, and software.

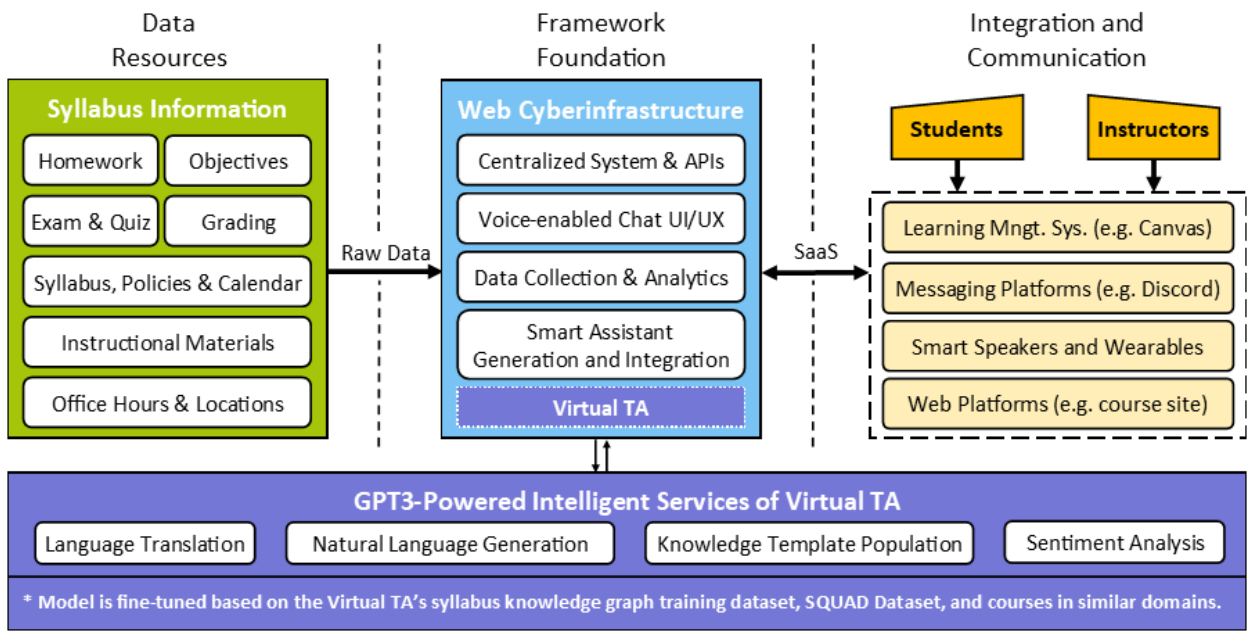

Figure 1. System architecture and components of VirtualTA

### 2.3.1. Intelligent Services

Course Knowledge Model Generation: In order to power the VirtualTA, the raw syllabus document needs to be parsed to create a knowledge model (Figure 2). The process for knowledge model generation entails utilizing GPT-3 to attempt to find relevant snippets out of unstructured text by using the competency questions provided in Table A.1. The retrieved information for each syllabus element is curated and stored in a JSON file. Upon post-processing and validation, the resulting knowledge model is used to fine-tune the model for question answering.

Pre-Processing: Once the syllabus file is parsed, the generated data is split into smaller pieces or documents to lower the cost to use the GPT-3 model and lower the latency time. The data was initially divided into 2000 characters, but this led to increased API request costs. We created our final version of the code to divide the syllabus content into documents with 200 characters without compromising accuracy or the model's affordability. It was made sure that the split does not divide a word into different syllables.

Post-processing: When the extracted text from the syllabus does not contain the information the question was intended for, the GPT-3 model may sometimes return irrelevant snippets to the asked question. In some cases, the model can return partial answers as well; partial in the sense that the response has the correct information, yet it is not complete (e.g., returns the information of 3 TAs out of 5 total TAs listed on the course description). To address and resolve these edge cases, the instructors (e.g., teaching assistant, faculty) are provided with the initial draft of the automatically populated knowledge graph and validate the information or modify as needed before the graph can be fed to the model for question answering. This is a one-time process, where the instructor(s) or TAs can go through the template at the start of the semester, to check the proposed answers by the model, and modify the knowledge base with accurate information. Throughout the semester, this workflow can be repeated as needed if major changes occur in the syllabus.

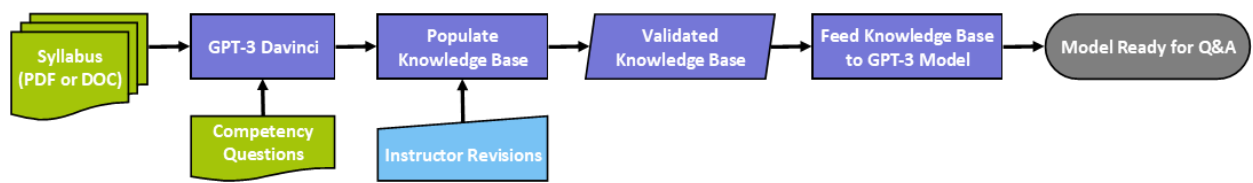

Figure 2. Knowledge base population process with instructor revision

Question Answering (QA): The questions answering process relies on the provided course knowledge model and two models to understand the intent, map it to the requested resource, and produce a natural language response in the form of a to-the-point and concise answer. The question-answering framework from GPT-3 by OpenAI works in two parts. The first part of the QA process is the search model, which is used to search the provided documents. This model then lists the most applicable documents to answer the given question. For this we created a fine-tuned model rather than using the models provided by OpenAI. Our fine-tuned model is trained on the Stanford Question Answering Dataset (SQuAD) dataset. Both the training and validation datasets have 1,314 entries each for our fine-tuned model. The second part of the QA process is the completion model, which is a built-in model provided by OpenAI named "text-davinci-002". Davinci is the most capable model family provided by OpenAI. This model is good at understanding content, summarization and content generation. Completion model is used to generate the answer from the documents provided by the Search Model.

As a way to expand the accessibility as well as the human-like and empathetic interactions of the system in a way that it can seamlessly serve with an approachable persona, several enhancements were devised and implemented. The question-answering system operates in a variety of languages. This was accomplished through the use of GPT-3's language translation capabilities. Students can ask a question in any language supported by GPT-3, which we then translate to English, send to VirtualTA, receive an answer in English, and finally translate back to the language the question was asked in. VirtualTA can be tailored to the demands of the students by fine-tuning the model using the questions asked by the students and answers provided by the model. This customization can be done at the domain or course level. The sentiment of a student's question can be analyzed by VirtualTA, and if negative emotions or stress are identified, the system will give positive comments or optimistic messages to lighten the situation and point them towards appropriate available resources.

### 2.3.2. Framework Integration

The framework is founded upon a centralized web-based cyberinfrastructure for data acquisition, training deep learning models, storage and processing of course-specific information, as well as to host the generated chatbots for use in communication channels. The cyberinfrastructure entails an NGINX web server, NodeJS-based backend logic, a PostgreSQL database, accompanied with caching, and user and course management mechanisms. The core intelligent assistant is created based upon the Service-Oriented Architecture, allowing its plug-and-play integration into any web platform with webhooks. While the system can be integrated into numerous channels (e.g., augmented and virtual reality applications, automated workflows, learning management systems), several integrations have been realized as part of this paper to showcase its utility.

<u>Web-based Conversational Interface:</u> To make asking questions and receiving responses easier, a web-based chatbot application user interface (UI) has been developed. The UI designed by Palace C was modified for this development using standard JS. Through the API we developed, VirtualTA's replies may be retrieved. The user sees this response on the chatbot provided by VirtualTA. This may be included into any web-based chatbot by using the API we established to receive an answer and utilize it as the chatbot's response. This procedure enables VirtualTA's functionality to be incorporated to any web-based conversational bot.

<u>Social Platforms:</u> A Discord bot is created to allow students to include the VirtualTA into workspaces they already use for specific courses to facilitate easy access to pertinent information. The availability of VirtualTA into social messaging platforms students already utilize allows for easy adoption of the system as well as an organic and friendly interaction.

<u>Smart Apps and Devices:</u> VirtualTA is integrated to work with Google Assistant. We have created an API to return an answer, when asked a question regarding a course. We have used Google DialogFlow to integrate the VirtualTA as a third-party action on Google Assistant. Students can access VirtualTA using Google Assistant on their mobile phones, smart home devices, Android TV, Android Auto and smart watches. This integration has been deployed in the test environment on this platform and screenshots of these implementations are shared in section 3 below.

### 2.4. Case Study Design

In order to establish the accuracy and performance of the VirtualTA, an assessment was conducted by collecting 112 syllabus files from a variety of institutions and domains, including

Engineering, Math, Physics, History, Computer Science, English, Art, Business, Philosophy, Arabic, Anthropology, Accounting, Chemistry, Music, and Economics. We removed 12 of these files because the syllabi were in image format and text extraction from the image could hinder the benchmark of VirtualTA's capabilities. Hence, a case study is designed upon 100 syllabi and in two phases to assess the performance for (1) extracting data from syllabi and (2) mapping user questions to the extracted syllabi data.

### 2.4.1. Phase 1 - Knowledge Extraction

We chose 38 files from the 100 syllabus files we collected. We asked the VirtualTA 36 questions we chose based on main categories for every syllabus file. The goal of this is to measure the accuracy of the bot on the frequently asked questions. We had three parameters we are collecting from this study: number of questions answered correctly, number of questions answered incorrectly and number of questions partially answered. These parameters could be seen below in the template that is created for one of the courses and is in JSONL format.

**Before Edits**

{"QUESTION":"What is the name of the course?","ANSWER":"BUS 100","isTrue":"Change this to TRUE or FALSE or PARTIAL"}
{"QUESTION":"What is the course number?","ANSWER":"The course number is BUS 100.","isTrue":"Change this to TRUE or FALSE or PARTIAL"}
{"QUESTION":"How many credit hours is this course worth?","ANSWER":"This course is worth 3 credit hours.","isTrue":"Change this to TRUE or FALSE or PARTIAL"}

**After Edits**

{"QUESTION":"What is the name of the course?","ANSWER":"Introduction to Business","isTrue":"FALSE"}
{"QUESTION":"What is the course number?","ANSWER":"The course number is BUS 100.","isTrue":"TRUE"}
{"QUESTION":"How many credit hours is this course worth?","ANSWER":"This course is worth 3 credit hours.","isTrue":"TRUE"}

Once we collect the answers from the bot on all the 36 questions for the syllabus file, we store these results in a JSONL file format. In the file we have three fields including question, answer and isTrue flag. The question field contains the question asked to the bot, the answer field contains the answer we got from the bot and the isTrue is whether the given answer is correct or incorrect. We manually went through all the 38 JSONL files and checked the answers with the actual syllabus file and change the isTrue field to "TRUE" if the answer given by the bot is correct, "FALSE" if the answer given by the bot is incorrect and "PARTIAL" if the answer given by the bot is partially correct. When an answer is false/incorrect, in addition to making the isTrue "FALSE", we also change the answer to the correct version/information. These manual corrections were done to use this information for our second phase of testing. The manual corrections explained above could be seen in Table 2 (After Edits), this figure shows the corrections made and isTrue field set to either "FALSE","TRUE" or "PARTIAL" and these changes have been highlighted in green.

### 2.4.2. *Phase 2 -* **Question Answering**

We use the manually corrected templates created in Phase 1. Here in this phase of testing we increase the questions asked from 36 to 70. This was done using text augmentation to test the model's question and answering performance on different question asking techniques or structures. Each question has at least one other variation except for two questions. The rationale for these two questions, "How do I submit my assignments?" and "when is the final exam?", left out from data augmentation is because we couldn't discover a sensible and good approach to supplement or augment these queries. We had three parameters collected during this study: number of questions answered correctly, number of questions answered incorrectly, and number of questions partially answered.

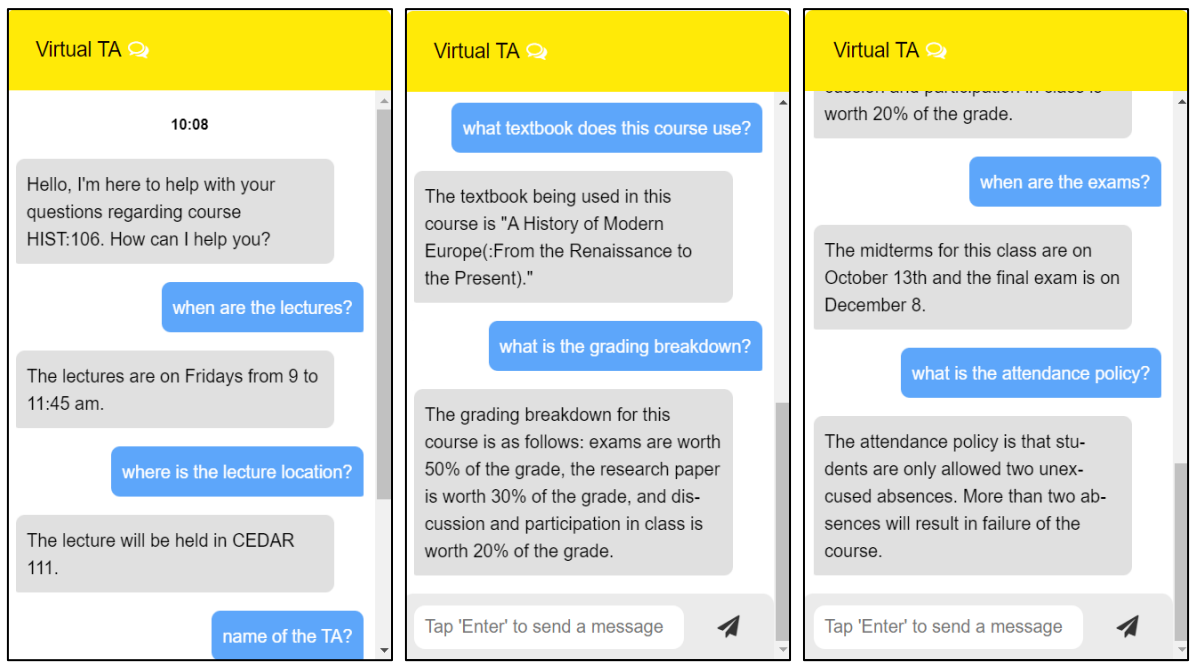

Figure 3. Web based chatbot user interface with questions and answers

Once we collect the answers from the bot on all the 70 questions for the template file created in Phase 1, we store these results in a JSONL file format. In the file, we have three fields including question, answer and isTrue flag. The question field contains the question asked to the bot, the answer field contains the answer we got from the bot and the isTrue field is whether the given answer is correct or incorrect. We manually went through all the 38 JSONL files and checked the answers with the actual syllabus file and change the isTrue field to "TRUE" if the answer given by the bot is correct, "FALSE" if the answer given by the bot is incorrect and "PARTIAL" if the answer given by the bot is partially correct.

## 3. Results and Discussion
### 3.1. Communication Channels

The user interface for the web platform shown in Figure 3 has been adapted from Palace, 2021. Figure 3 shown below is a web platform designed using vanilla JavaScript. Figure 3 shows select Competency Questions asked to the model and the answers returned by the model for a history course.

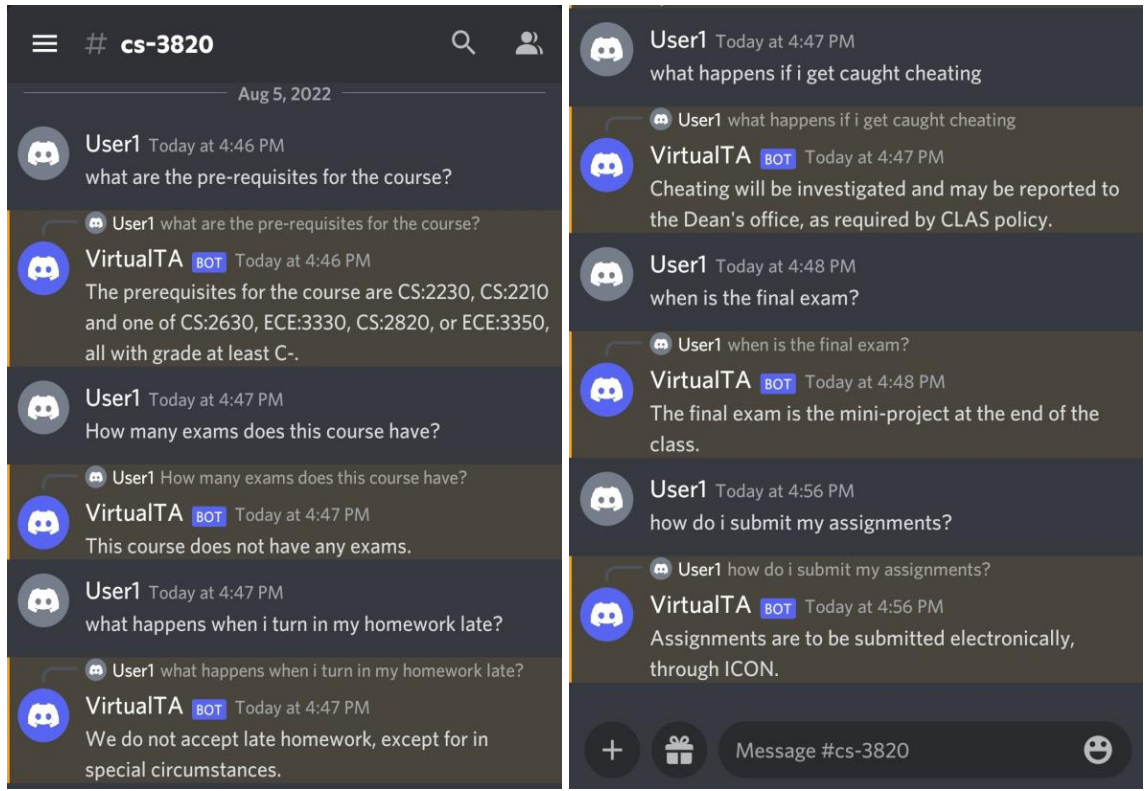

Figure 4. Integration of VirtualTA with Discord social media application

Figure 4 show the third-party integration of VirtualTA model and in this case the third-party application is Discord. In the figure the questions asked to the model and the responses given by the model for a STEM course specifically a CS course is provided.

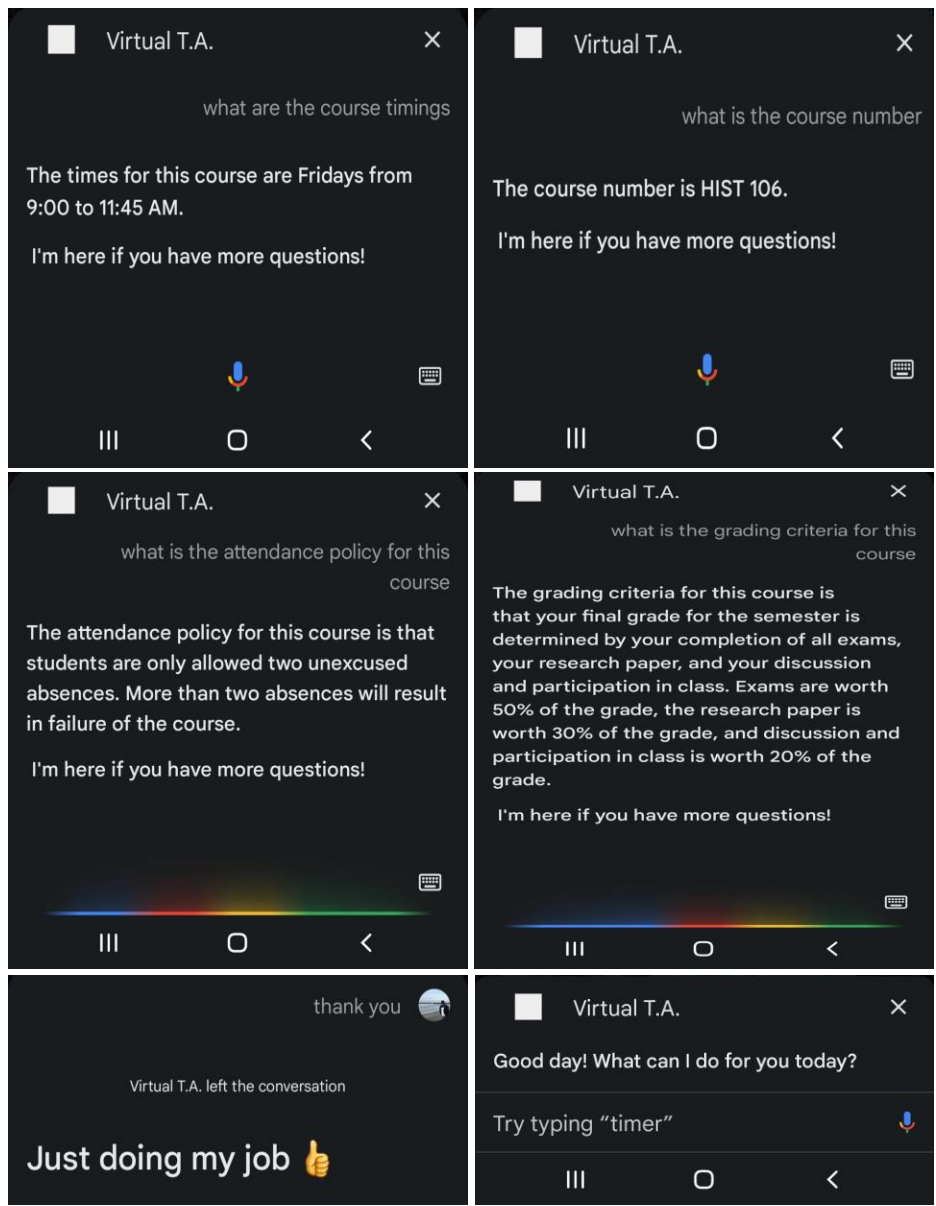

Figure 5. Integration of Virtual TA with Google Assistant application

Figure 5 shows the integration of VirtualTA with Google Assistant. The questions are asked to VirtualTA using voice. The command "talk to Virtual T.A." is needed to connect Google Assistant to the third-party action VirtualTA. These figures show the questions asked and the answers returned by VirtualTA for a history course.

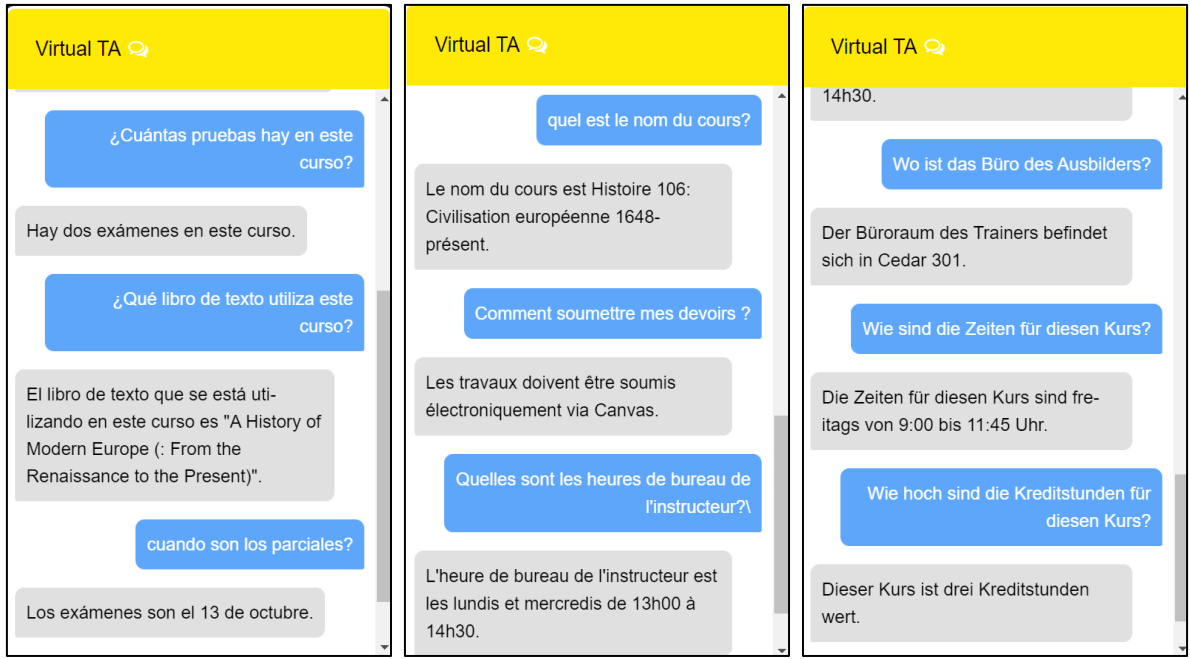

Figure 6. VirtualTA language translation capabilities

Figure 6 illustrates the capability of language translation in VirtualTA. The VirtualTA support languages for Spanish, French, and German are shown in (A), (B), and (C), respectively. Any language that GPT-3 supports can be used by the user to ask a query, and VirtualTA will respond in that language.

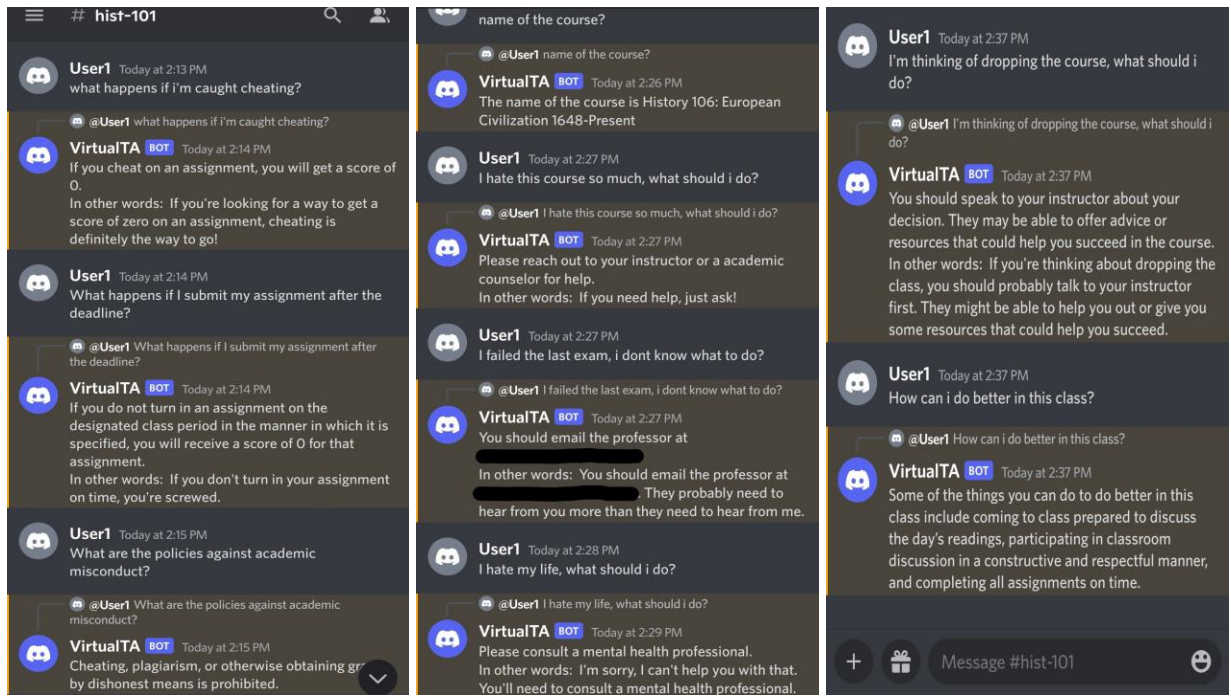

Figure 7. VirtualTA sentiment analysis for a history course

The possibilities of VirtualTA's sentiment analysis are displayed in Figure 7. The model responds with the standard response and also in a humorous or witty way to lighten the situation if it determines that the user is asking a question with a negative emotion or feeling. Figure 7 show some examples of sentiment analysis in VirtualTA. Private information, including the instructor's email address, has been obscured in Figure 7 (center).

### 3.2. Performance Evaluation

To quantify the model's effectiveness, precision (Equation 1), recall (Equation 2), and f1-score (Equation 3) metrics have been selected for this imbalanced classification problem with multiple classes as formulated below (Sokolova and Lapalme, 2009). *n value* in the equations below represents the number of different questions in the FAQ (i.e., classes). For computing results using equations 1-3 (Sokolova & Lapalme, 2009), we have used the criteria listed below and computed two sets of results where one includes "PARTIAL" as correct/TruePositive and the other does not consider "PARTIAL" as correct/TruePositive.

$$\text{Precision (multiclass)} = \frac{\sum_0^n TP}{\sum_0^n TP + \sum_0^n FP} \quad (1)$$

$$\text{Recall (multiclass)} = \frac{\sum_0^n TP}{\sum_0^n TP + \sum_0^n FN} \quad (2)$$

$$\text{f1 - score (multiclass)} = \frac{2 \times precision \times recall}{precision + recall} \quad (3)$$

The aim of the testing phase is to optimize the precision and recall values to build an accurate and complete system, however a trade-off evaluation is necessary (He and Ma, 2013). Depending on the use case, it may be necessary to maximize precision to ensure an accurate answer is provided or to maximize recall to map as many questions as possible while limited sacrifice of the accuracy. For this use case, we tried to maximize the precision value for the model.

For each of the 38 syllabus files utilized in the testing phase, these performance values were calculated and analyzed. The results in Table 3 from the knowledge extraction sub-section and Table 4 from question answering sub-section represent the average accuracy values across all 38 files. VirtualTA prioritizes accuracy of the responses over giving an answer. We want to provide the most accurate results to the students. It is better to respond with "Answer not found" than to give an incorrect answer. Especially where the incorrect answer could misinform a student and can lead to them missing office hours or homework deadlines. When the model is unsure of the answer or is unable to locate pertinent documents, it responds to the user as "Response not found," which prevents it from providing the user an incorrect answer and allows them to double-check the answer outside of VirtualTA with instructor or TA.

### 3.2.1. Knowledge Extraction

These values shown in Table 2 are calculated by computing the regular accuracy which is correct answers over the total number of questions. The common problems faced by the model is in the "Teaching Assistant Information" section, and this could be due of many reasons, and we identified some of these cases as the reason for lower accuracy in our case study as follows: (1) when a course or syllabus has multiple TAs (Teaching Assistants), the model is failing to detect all the TAs from the text provided, (2) when a course doesn't have a TA listed; the model fills in the TA questions with the instructor's information. For instance: when there is no TA for the course and user asks, "when are the TA's office hours?" the model replies back with the instructor's office hours, (3) the formatting of some syllabuses we tested on were really confusing or messy. This resulted in the model missing simple questions such as the course number or course name. In calculating these results discussed in this section, we considered all of the above cases (1-3) as Incorrect/False.

Table 2. Accuracy for Phase 1 testing results

| Category name | Number of questions per category | Accuracy | Accuracy with partial content |
|---|---|---|---|
| Course Information | 6 | 63.2% | 69.7% |
| Faculty Information | 4 | 60.5% | 73.0% |
| TA information | 4 | 49.3% | 57.2% |
| Course Goals | 2 | 90.8% | 94.7% |
| Course Calendar | 3 | 90.4% | 92.1% |
| Attendance | 2 | 93.4% | 97.4% |
| Grading | 4 | 63.8% | 71.0% |
| Instructional Materials | 2 | 92.1% | 96.0% |
| Policies | 9 | 95.0% | 95.6% |
| Overall | 36 | 76.5% | 81.6% |

Table 3. Performance metrics for Phase 1 testing

| Performance Metrics | Without Partial | Includes Partial |
|---|---|---|
| Recall | 0.64 | 0.69 |
| Precision | 0.79 | 0.81 |
| f1-score | 0.71 | 0.74 |

Table 4. Performance metrics for Phase 2 testing

| Performance Metrics | Without Partial | Includes Partial |
|---|---|---|
| Recall | 0.87 | 0.88 |
| Precision | 0.96 | 0.96 |
| f1-score | 0.91 | 0.92 |

### 3.2.2. Question Answering

The values shown in Table 5 are calculated by computing the regular accuracy which is correct answers over the total number of questions. The common problems faced by the model in this phase is in the "Course Information" section, and this could be because of many edge cases, and we identified two main cases as the reason for lower accuracy in our case study as follows: (1) when the number of credit hours are not given in the context, the model tries to calculate the credit hours based on the number of lectures per week; (2) there is a small chance where the model can give different answers to similar questions, this could be based on the style of questioning. In calculating these results discussed in this section, we considered all of these above cases as Incorrect/False.

Table 5. Accuracy values for Phase 2 testing

| Category name | Number of questions per category | Accuracy | Accuracy with partial content |
|---|---|---|---|
| Course Information | 12 | 82.2% | 83.6% |
| Faculty Information | 8 | 97.4% | 99.0% |
| TA information | 8 | 98.7% | 98.7% |
| Course Goals | 4 | 98.0% | 99.3% |
| Course Calendar | 5 | 99.5% | 100% |
| Attendance | 4 | 98.7% | 100% |
| Grading | 7 | 95.5% | 98.5% |
| Instructional Materials | 4 | 96.0% | 97.4% |
| Policies | 18 | 99.0% | 99.6% |
| Overall | 70 | 95.3% | 96.5% |

### 4. Conclusion and Future Work

In this research, we designed an automated system for answering logistical questions in online course discussion boards, third party applications or educational platforms and highlighted how it can aid in the development of virtual teaching assistants. Specifically, the project's aims include enhancing course content quality and individualized student advising by delegating the time-consuming repetitive duties of instructors to virtual assistants and mitigating inequality among students in accessing knowledge to narrow retention and graduation gaps. This document outlines the entire procedure and displays the VirtualTA architecture. Through this architecture, VirtualTA can be integrated with third-party applications to enable access from a variety of intermediaries, such as web-based systems, agent-based bots (such as Microsoft Skype and Facebook Messenger), smartphone applications (such as smart assistants), and automated web workflows (e.g., IFTTT, MS Flow). Users will find it simple to access VirtualTA through any communication channel that is familiar to them and that they feel comfortable using. Additionally, it enables any number of users enrolled in the course to access the system. We want to expand upon our existing approach to include course content in addition to the syllabus or administrative information.

Furthermore, several future studies are possible including: (1) a case study with students for a semester long course in multiple fields/departments, (2) integrating the VirtualTA with learning management systems, (3) creating course content assistance and search, quiz and flash card mechanisms, (4) integration to other main-stream communication channels, (5) personalized communication to the pace and language for the student's level of understanding, (6) improving the system accuracy and performance by fine tuning the model on bigger datasets, (7) developing methods to understand different question asking techniques, and (8) integrating necessary course information directly from learning management systems.

**Acknowledgement:** This material is based upon work supported by the National Science Foundation under Grant No. #2137891 and #2230710.

## 6. Appendix

Table A.1. List of competency questions for the syllabus knowledge graph

| Category | Syllabus Element | Competency Questions |
|---|---|---|
| Course Information | Course Name | What is the name of the course? |
| | | What is the course's title? |
| | | What is the course's name? |
| | Course Number | What is the course number? |
| | | What is the number of the course? |
| | | What is the course ID? |
| | Credit Hours | How many credit hours is this course worth? |
| | | What are the credit hours for this course? |
| | | What is the credit hour value of this course? |
| | Location and Class Times | Where is the location of the lecture? |
| | | Where will the lecture be held? |

| | | | |
|---|---|---|---|
| | | What will be the lectures location? | |
| | | What times are the lectures? | |
| | | When do the lectures take place? | |
| | Prerequisites/Co-requisites | Does this course have any prerequisites/corequisites? | |
| | | Are there any prerequisites or corequisites for this course? | |
| | | Is there a pre-requisite or a co-requisite for this course? | |
| Faculty Information | Name | What is the name of the instructor/professor? | |
| | | What is the name of the professor or instructor? | |
| | | What is the professor's name? | |
| | Contact Information | How can I contact the instructor? | |
| | | What is the best way for me to get in touch with the instructor? | |
| | | What is the contact information of the Instructor? | |
| | Office Location | Where is the office of the instructor? | |
| | | Where can I find the instructor's office? | |
| | | What is the location of the instructor's office? | |
| | Office Hours | When are the instructor's office hours? | |
| | | What are the instructor's office hours? | |
| | | How long is the instructor's office open? | |
| TA/Teaching Assistant Information | Name | What is the name of the TA/Teaching Assistant? | |
| | | What is the TA's/Teaching Assistant's name? | |
| | | What is the name of the TA? | |
| | Contact Information | How can I contact the TA? | |
| | | What is the best way for me to get in touch with the TA? | |
| | | What is the TA's email? | |
| | Office Location | Where is the office of the TA? | |
| | | Where can I find the TA's office? | |
| | | What is the location of the TA's office? | |
| | Office Hours | When are the TA's office hours? | |
| | | What are the TA's office hours? | |
| | | How long is the TA's office open? | |
| Course Goals and/or Objectives | Course Objectives | What are the course objectives? | |
| | | What are the goals of the course? | |
| | | What are the objectives of this course? | |

| | Expectations from the course | What are the expectations from the course? |
|---|---|---|
| | | What can you anticipate from the course? |
| | | What are the course's objectives? |
| Course calendar | Due dates | When are Assignments due? |
| | | When are the deadlines for assignments? |
| | | When must assignments be completed? |
| | Assignment dates | When are assignments released? |
| | | When do assignments become available? |
| | | When will assignments be made available? |
| Attendance and classroom behavior | Attendance policy | What is the attendance policy for this course? |
| | | What is the course's attendance policy? |
| | | What is this class's attendance policy? |
| | Expected Classroom behavior | Is there any expected classroom behavior? |
| | | Is there any standard classroom conduct? |
| | | Is there any anticipated behavior in the classroom? |
| Grading and Assignments | Grading Criteria | What is the grading criteria for this course? |
| | | What is the grading breakdown? |
| | | What are the grading guidelines for this course? |
| | Tentative Exam Schedule | How many exams does this course have? |
| | | When are the exams? |
| | | When will the exams be held? |
| | | When is the final exam? |
| Instructional Materials | Textbooks | What textbook does this course use? |
| | | What textbook is being used in this course? |
| | | What textbook is used in this class? |
| | Other required materials for the course | What are the other materials this course uses? |
| | | What additional materials does this course use? |
| | | What additional materials does this course make use of? |
| Policies | Late Assignments | What happens when I turn in my assignment late? |
| | | What happens if I submit my assignment after the deadline? |
| | | What happens if I miss the deadline for submitting my assignment? |
| | Academic dishonesty | What is the cheating policy? |
| | | What are the policies on Academic dishonesty? |

|  |  | What are the policies against academic misconduct? |
|---|---|---|
|  | Disability statement | What are the accommodations provided for a student with a disability? |
|  |  | What kind of accommodations are made for a student who has a disability? |
|  | Freedom of speech | How does this class accommodate Free speech and Expression? |
|  |  | How does this class allow for free speech and expression? |
|  | Makeup Policy | What is the makeup policy for assignments? |
|  |  | What is the policy on assignment makeup? |
|  | Mental Health Resources | Are there any resources for mental health? |
|  |  | Is there any information about mental health? |
|  |  | Are there any mental health resources? |
|  | Absences for Religious Holy Days | How do absences for holy days be considered? |
|  |  | How are absences from holy days taken into account? |